\newcommand{\PaperTitle}{Initial Evidence of Elevated Reconnaissance Attacks Against Nodes in P2P Overlay Networks}

\newcommand{\PaperNumber}{26}

\documentclass[10pt,sigconf,letterpaper,nonacm]{acmart}

\usepackage{enumitem}
\usepackage{graphicx}
\usepackage{float}
\usepackage{textcomp}
\usepackage{xcolor}
\usepackage{hyperref}
\usepackage{caption,subcaption}
\usepackage{pifont} %
\usepackage{array} %
\usepackage{tabularx} %
\usepackage{xurl}
\usepackage{hyperref}
\usepackage{multicol}
\usepackage[capitalise,noabbrev]{cleveref} 
\crefname{appsec}{Appendix}{Appendices} 
\usepackage{setspace}

\newcommand\blfootnote[1]{%
  \begingroup
  \renewcommand\thefootnote{}\footnote{#1}%
  \addtocounter{footnote}{-1}%
  \endgroup
}

\begin{document}

\title{\PaperTitle}
\settopmatter{printfolios=true}

\author{Scott Seidenberger}
\affiliation{%
  \institution{University of Oklahoma}
  \city{Norman}
  \state{OK}
  \country{USA}
}
\email{seidenberger@ou.edu}

\author{Anindya Maiti}
\affiliation{%
  \institution{University of Oklahoma}
  \city{Norman}
  \state{OK}
  \country{USA}
}
\email{am@ou.edu}

\begin{abstract}
We hypothesize that peer-to-peer (P2P) overlay network nodes can be attractive to attackers due to their visibility, sustained uptime, and resource potential. Towards validating this hypothesis, we investigate the state of active reconnaissance attacks on Ethereum P2P network nodes by deploying a series of honeypots alongside actual Ethereum nodes across globally distributed vantage points. We find that Ethereum nodes experience not only increased attacks, but also specific types of attacks targeting particular ports and services. Furthermore, we find evidence that the threat assessment on our nodes is applicable to the wider P2P network by having performed port scans on other reachable peers. Our findings provide insights into potential mitigation strategies to improve the security of the P2P networking layer.
\blfootnote{Partly supported by Ethereum Foundation grants FY24-1546 and FY24-1547.}

\end{abstract}

\maketitle

\section{Introduction}

Cyber threats are rapidly evolving, with attackers increasingly targeting Internet-connected devices~\cite{butun2019security,haddadpajouh2021survey}. These devices are attractive for malicious activities, such as deploying malware, acting as command-and-control (C2) intermediaries, or serving as botnet components. P2P networks, foundational for decentralized applications, file sharing, and blockchain, make their nodes accessible through open routing information~\cite{rpcporthoneypot}. While essential for network functionality, this transparency inadvertently exposes nodes to significant risks~\cite{attackactivitieshoneypotethereum}. However, systematic threat assessments focusing on P2P nodes' security remain scarce~\cite{smartindustrycloudhoneypot}.

This work is an initial study into elevated reconnaissance activities targeting nodes in P2P overlay networks, focusing on capturing and understanding attack patterns. To investigate these dynamics, we introduce Ethereum as a case study due to its prominent role as one of the largest and most active P2P networks. Ethereum supports over 8,000 publicly accessible nodes globally~\cite{grandjean2023ethereumproofofstakeconsensuslayer}, which regularly broadcast routing and metadata information to support decentralized network functionalities. This open architecture and high visibility can make Ethereum nodes especially attractive targets for malicious actors, underscoring the importance of quantifying these risks and identifying specific vulnerabilities. 

Our approach combines a honeynet with actual Ethereum nodes, deployed across geographically distributed vantage points (VPs), to capture a wide range of attack vectors. This deployment enables us to document %
evidence of heightened reconnaissance specifically targeting Ethereum nodes in P2P networks. By focusing on Ethereum, we aim to provide a comprehensive assessment of potential threats that may similarly affect other decentralized networks.
Our analysis includes a longitudinal assessment of threats targeting Ethereum nodes and their impact on node performance. We find that nodes experience not only increased attacks but also specific types of attacks targeting particular ports and processes. Additionally, we examine the network-level distribution of attack sources, finding they diverge from the networks hosting these Ethereum nodes, suggesting that attacks primarily originate from different regions rather than being self-originated within the P2P network itself~\cite{attackactivitieshoneypotethereum}.

Our contributions in this work are as follows:
\begin{itemize}[leftmargin=*]
    \item \textbf{Measurement:} We deployed a data collection infrastructure on AWS across five geographic regions, including honeypots and beacon nodes, to analyze P2P interactions and connection stability.
    
    \item \textbf{Analysis of Attack Patterns:} Using statistical models accounting for geographic variations, we compared attack patterns on Ethereum nodes and control nodes. Ethereum nodes faced significantly more attacks, including targeted URI requests for sensitive files and cloud metadata.
    
    \item \textbf{Assessment of Attack Strategies:} SSH login attempts were more frequent on Ethereum nodes, indicating focused attack strategies with reused passwords. Scanning TCP ports revealed many Ethereum nodes had open SSH and RPC ports, increasing their attack surface and highlighting the need for strong security measures.

\end{itemize}

These findings highlight the need to conduct similar measurement studies on other P2P networks to better understand and mitigate potential security vulnerabilities across decentralized systems. Based on our observations, we recommend network-layer enhancements, such as privacy-preserving routing protocols, and advocate for collaborative threat intelligence sharing. This work serves as a foundation for future studies on securing critical infrastructure in decentralized networks.

\section{Background and Related Works}

Deploying honeypots on cloud infrastructure enables wide-scale cyber threat intelligence collection~\cite{correlationofcyberthreatintel,smartindustrycloudhoneypot}, aiding researchers and security professionals in monitoring new threats. Enterprises also deploy honeypots internally to complement intrusion detection systems (IDS)~\cite{enterprisecloudhoneypot}. However, scaling data collection in large honeypot deployments, or honeynets, presents operational challenges~\cite{alata2006lessons,barford2010employing}. Leveraging prior work, we have deployed a horizontally scalable approach for collecting data from our honeynet alongside Ethereum node metrics and network session data~\cite{honeypotusingelk}. Complementary to honeypots~\cite{soro2023enlightening}, darknet sensors monitor unused IP address spaces for unsolicited traffic, offering an alternative perspective in internet-scale monitoring~\cite{fachkha2015darknet}.

Our methodology extends single-protocol honeypots by utilizing meta-honeypots that handle non-standard services and ports~\cite{honeytrap,tpot,cowrie,glutton}, allowing us to collect a broader spectrum of attack vectors. 
Similar security challenges have been observed in Internet of Things (IoT) devices, which are always on, unattended, and share application-specific characteristics with P2P network nodes~\cite{zhang2022intrusion,wu2020grouptracer}. Attackers exploit these vulnerabilities to orchestrate large-scale attacks like the Mirai botnet's DDoS attacks~\cite{antonakakis2017understanding}. We hypothesize that P2P nodes, sharing these characteristics and being more resource-rich, are appealing new targets for attackers seeking persistent hosts.

In P2P overlay networks, the exploitation of public routing information is an inherent vulnerability~\cite{imamura2019network}. Studies have analyzed risks from de-anonymized routing data, exposing overlay networks to targeted attacks~\cite{neudecker2015simulation,bojja2017dandelion}. Specifically for Ethereum, measurement studies have examined its topology and P2P characteristics to understand the potential attack surface~\cite{kim2018measuring,gao2019topology,kiffer2021under,wang2021ethna,hu2022identifying}.
Previous work established this attack surface under Ethereum's proof-of-work (PoW) consensus model~\cite{chin2021analysis}, where nodes were tightly coupled to the network's economic layer, as miners had wallets associated with their nodes. Since Ethereum switched to proof-of-stake (PoS) in 2022, nodes are two degrees separated from the economic layer. %
Applying cloud-based honeypots to the Ethereum network, prior studies have shown evidence that attackers target vulnerable Ethereum nodes~\cite{attackactivitieshoneypotethereum}. Attackers often locate these targets by extracting public routing information from Ethereum's Kademlia distributed hash table (DHT). This work examines a single vulnerable port--the RPC port associated with the Ethereum client software--to execute malicious commands aimed at stealing Ether from the target node's wallet~\cite{rpcporthoneypot,stealingattackfirststep}. 

Our work complements and extends these studies by having honeypots listen on all available ports, not just the standard Ethereum RPC port. This broader approach enables us to explore threats arising from server misconfigurations or vulnerabilities beyond an exposed and insecure RPC port. Considering the shift to PoS consensus, the specific attacks identified in prior studies are now less practical. Therefore, it's increasingly important to understand the security of the servers running the nodes themselves, rather than focusing solely on the execution client's RPC port. To address these emerging security concerns, our experimental design tests the hypothesis that P2P overlay network nodes are lucrative targets for attackers seeking persistent hosts.

\section{Experimental Setup}

To test the hypothesis that P2P network nodes are being targeted, we deployed several mainnet Ethereum nodes and a diverse set of honeypots on globally distributed VPs for a two-month period. 

\subsection{Global Vantage Points}

We deployed 10 geographically diverse VPs via AWS EC2. The VP servers were assigned to a control group and an experimental group, with a pair of each deployed to five distinct geographic regions. On each, we ran a honeypot platform, T-Pot~\cite{tpot}, with a diverse set of honeypot containers with full port coverage at varying levels of interactivity, as well as the Arkime~\cite{arkime} network traffic monitoring service. The experimental group additionally ran Ethereum beacon nodes on mainnet, allowing for a direct comparison of honeypot data between the two groups. The EC2 instances are Ubuntu base images configured as in \cref{tab:group-comparison}; %
the AMIs (Amazon Machine Images) are made available as artifacts of the dataset collection. 

\begin{table}[H]
\centering \small
\caption{Comparison of Control and Experimental group server characteristics.}
\label{tab:group-comparison}
\begin{tabular}{|c|c|c|}
\hline
\textbf{Characteristic}        & \textbf{Control}             & \textbf{Experimental (Node)}                 \\ \hline
Disk Size (GB)                 & 128                                & 2500                                \\ \hline
Instance Type                  & m7/6i.large              & r7/6i.xlarge            \\ \hline
Software            & T-Pot, Arkime                      & LH, NM, T-Pot, Arkime \\ \hline
\end{tabular}
\end{table}

One of each server type was deployed across the following AWS regions to maximize global coverage:

\begin{itemize}%
    \item \textbf{NA (us-east-1)}: Virginia, US
    \item \textbf{EU (eu-central-1)}: Frankfurt, DE
    \item \textbf{ME (me-central-1)}: United Arab Emirates, AE
    \item \textbf{AP (ap-east-1)}: Hong Kong, HK
    \item \textbf{SA (sa-east-1)}: São Paulo, BR
\end{itemize}

\subsection{Ethereum Node Deployment}

An Ethereum beacon node combines an Ethereum consensus client and execution client that communicate with other nodes to form the network~\cite{ethereum_nodes_clients}. Nodes are responsible for enforcing the protocol, serving block data, and acting as entry points for validator clients to propose and attest to new blocks. The consensus client--in our case, Lighthouse (LH)~\cite{lighthouse}--communicates with peers over \texttt{discv5} and \texttt{libp2p} to determine the current, canonical head of the chain. The execution client--here, Nethermind (NM)~\cite{nethermind}--uses \texttt{discv4} and \texttt{devp2p} to share transactions and compute the current global state of the Ethereum Virtual Machine (EVM). 

When deployed together, these two clients form a beacon node. We deployed our LH and NM clients with default settings, except for the advertised P2P ports being 64000 (consensus) and 64333 (execution) to avoid conflicts with the honeypots. In our setup, the validator client is simulated by the Lighthouse Attestation Simulator~\cite{lighthouse_validator_monitoring}, a built-in function that produces simulated attestations each epoch to measure the performance of the node as if it were a validator on the mainnet. This data is key to understanding node performance and health without funding actual validators (currently requiring approximately USD 100k per validator). Each client exposes standardized metrics in order to monitor the client's behavior in detail.

\subsection{Honeypot Deployment}

The T-Pot platform manages multiple containerized, open-source honeypots at varying interaction levels. We deployed honeypots covering ports 1-63999 with the firewall open to all inbound traffic on these ports, while ports 64000 and above were reserved for management and the legitimate Ethereum node P2P ports. All interactions over the collection period are logged and stored. A full list of the honeypots used, along with their targeted services and ports are in \cref{tab:honeypot-containers} in the Appendix. Ensuring a broad coverage of services and ports is key to establishing robust baselines for later analysis. For this paper, we define an ``attack'' as a logged interaction with one of the honeypot containers~\cite{ilg2023survey}. 

\section{Findings}

\begin{figure}[t]
    \centering
    \includegraphics[width=\linewidth]{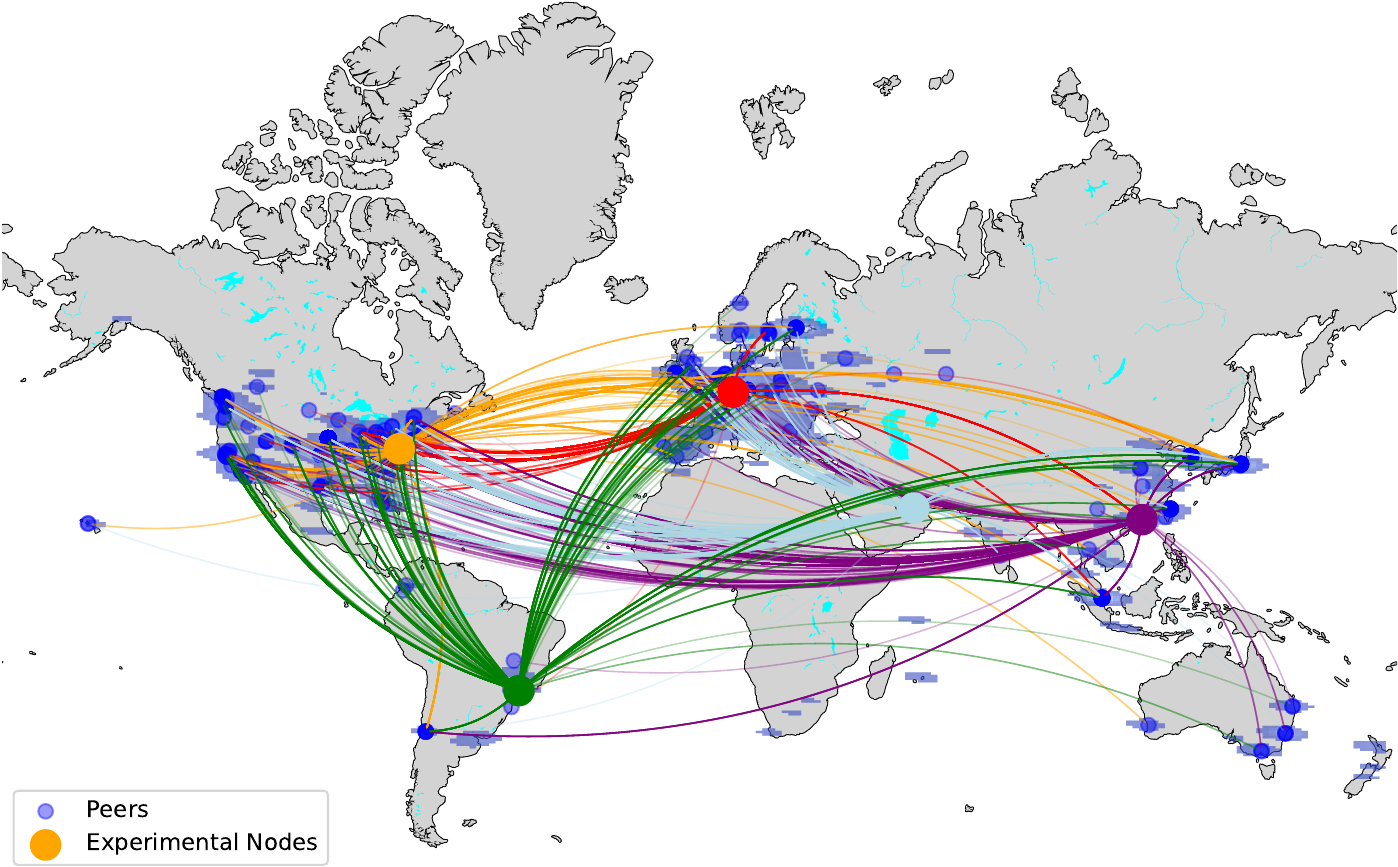}
    \caption{Top 10\% of peers by data volume with underlying heatmap of all scanned Ethereum nodes.}
    \label{fig:peers}
\end{figure}

After the 2-month collection period, we recorded 130.9 million attacks from 12.5 million unique source IPs on the honeypots across the 10 VPs. From the 5 beacon nodes, we collected 45.4 million P2P TCP Sessions to use in our analysis. \cref{fig:peers} illustrates the global peer coverage of our nodes using the P2P session data.

\begin{figure}[t]
    \centering
    \includegraphics[width=\linewidth]{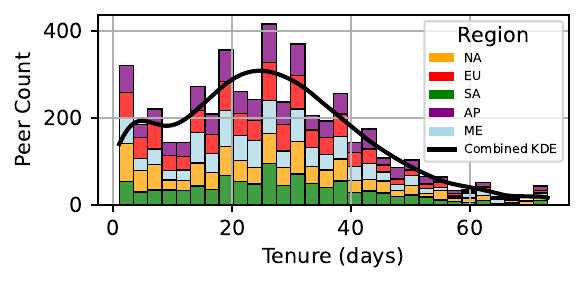}
    \caption{Peer tenure distribution, measured as non-consecutive connection days for peers grouped by unique IPs across regions. The distribution highlights stable, recurrent connections, suggesting P2P nodes provide a platform for long-term adversarial access.}
    \label{fig:peer-tenure}
\end{figure}

\textbf{Peer Tenure.} We first empirically established the tenure distribution for peers connected to the beacon nodes throughout the collection period. To quantify tenure, we grouped the P2P sessions of each peer by unique IP address and calculated the total number of distinct days each peer maintained a connection to a beacon node. By counting non-consecutive connection days, we captured the persistence of each peer without penalizing short gaps in connectivity that naturally occur in P2P networks. We generated a distribution of peer tenures of the top 1000 peers by data volume for each region, shown in \cref{fig:peer-tenure}, allowing us to confirm patterns across different geographical contexts. With a consistent average tenure ($\mu=26.46, n=5000$) across regions, our findings indicate a significant proportion of peers maintained stable, recurrent connections to the beacon nodes. This stability suggests that P2P nodes provide a persistent and reliable environment for adversaries seeking long-term access.

\subsection{Ethereum Nodes are Targeted}

To determine whether our Ethereum nodes are attacked more frequently than our control nodes, we needed to account for both temporal and geographic variations that could influence the number of attacks. We employed a generalized linear mixed model (GLMM) with a log-transformed response variable which allowed us to include both fixed effects (experimental vs. control groups) and random effects (geographic regions). The log transformation was applied to stabilize the variance and address the skewness inherent in count data.%

The results of the GLMM, see \cref{tab:glmm_results}, indicate that the experimental group received significantly more attacks than the control group after accounting for geographic variation. The experimental group had a significantly higher log-transformed attack count ($\hat{\beta}_1 = 1.111$, $SE = 0.008$, $z = 147.289$, $p < 0.001$). %
Model diagnostics indicated successful convergence, with the random effects capturing variation attributable to to geographic differences. The variance estimate for the group effect ($\sigma^2 = 0.050$, $SE = 0.030$) indicates that the effect size of regional variance is small. This suggests that much of the observed variability can can be attributed to the fixed effect of the presence of the P2P node rather than random geographic differences.

These findings support our initial hypothesis that Ethereum nodes are specifically targeted with a higher volume of attacks compared to control nodes, even after accounting for both time and geographic differences. Now, we will explore the nature of those attacks by what ports and services are targeted most commonly.

\subsection{Specific Port and Process Targeting}

\begin{figure}[t]
    \centering
    \includegraphics[width=0.8\linewidth]{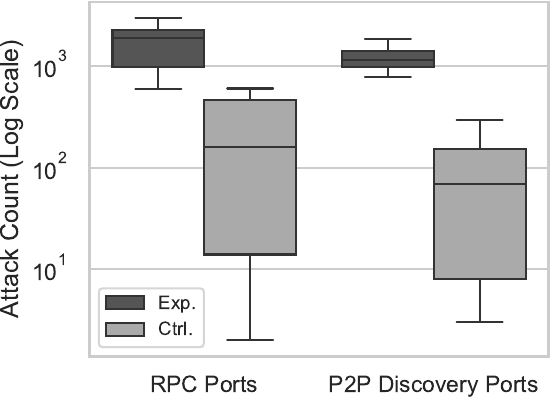}
    \caption{Boxplot highlighting increased reconnaissance against RPC and discovery ports of experimental group.}
    \label{fig:ports}
\end{figure}

\textbf{Port Targeting.} \Cref{fig:ports} confirms prior work that Ethereum RPC nodes are targeted, but contributes new findings that the ports are still targeted even under PoS consensus. Furthermore, we show new evidence that the P2P discovery ports are subject to reconnaissance, indicating that adversaries may be seeking vulnerable software to exploit in the node's networking stack, such as unpatched node software versions or other networking libraries.

\textbf{Credential Harvesting Requests.} URI request data captures attempts by external actors to access specific endpoints or files on the server via HTTP requests, revealing potential attack vectors. By comparing the URI requests made to the control and experimental groups, %
we can detect patterns of targeted behavior. %

\begin{figure}[t]
    \centering
    \includegraphics[width=0.95\linewidth]{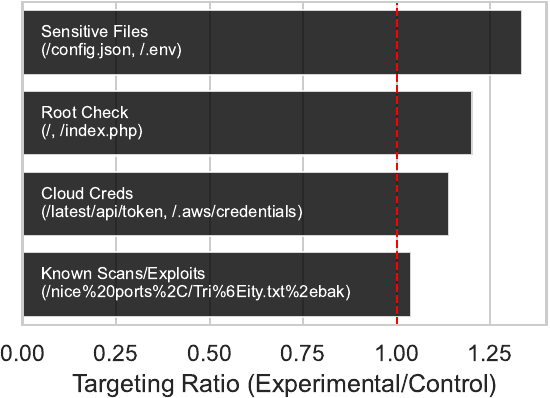}
    \caption{Significant categories of HTTP URI requests with request examples.}
    \label{fig:tgting_ratio}
\end{figure}

Our analysis reveals a concerning focus on accessing sensitive files and cloud metadata endpoints. Specifically, requests targeting files such as \texttt{.env} files and cloud-related URIs like \texttt{/latest/meta-data/iam/security-\linebreak credentials/} and \texttt{/.aws/credentials} suggest organized attempts to exploit misconfigurations of the infrastructure. These URIs often store environment variables, credentials, and API keys, potentially allowing attackers to obtain unauthorized access to cloud infrastructure or compromise the nodes. We group similar URI requests into categories and present significant differences in targeting ratio in \cref{fig:tgting_ratio}, where a targeting ratio of 1 would indicate equal attempts between the experimental and control groups. 

The recurring patterns in these requests point to a systematic attack strategy aimed at exploiting known vulnerabilities in Ethereum node deployments. The high frequency of requests for \texttt{.env} files and cloud-specific endpoints underscores the attackers’ focus on harvesting credentials or API keys to gain control over nodes or manipulate their behavior. %
It is important to note that node operators may run their services behind a reverse proxy, which leads to the use of these HTTP servers.

\textbf{Concentrated Authentication Attempts.} We derived Gini coefficients from the SSH login attempts to assess the distribution of attempted username and password combinations between control and experimental groups. The Lorenz curve of \cref{fig:lorenz} offers a visual representation of the inequality in distribution by plotting the cumulative share of user/pass combinations against the cumulative share of login attempts. By calculating the area between the Lorenz curve and the line of perfect equality, we derived the Gini coefficient, a numerical measure ranging from 0 (perfect equality) to 1 (perfect inequality), which quantifies the degree of concentration in the attempts.

The results reveal a significant difference in the concentration of login attempts between the two groups. The control group exhibited a Gini coefficient of 0.25, indicating a relatively even distribution of login attempts across a wide range of user/pass combinations. Conversely, the experimental group had a higher Gini coefficient of 0.43, meaning login attempts were more heavily concentrated on fewer combinations. This heightened concentration in the experimental group implies that attackers may be employing more focused or sophisticated strategies, possibly targeting specific credentials with higher probability of success.

While the password attempts had the same average length, the experimental group passwords had both lower average entropy ($p < .001$) and lower password reuse frequency. 

\subsection{Distribution of Attack Sources}

To determine whether inbound attacks originated from the same networks hosting nodes, we analyzed the distribution of attack sources and node locations at the Autonomous System (AS) level. Recognizing that background attack activity can vary significantly across different regions and network types-- and that cloud IPs receive orders of magnitude more traffic \cite{pauley2023dscope}--we adjusted for regional effects using control data. By subtracting the attack counts observed by control nodes from those of Ethereum nodes for each AS, we obtained adjusted attack counts that more accurately reflect attacks specifically targeting Ethereum nodes.

\begin{figure}[t]
    \centering
    \includegraphics[width=0.95\linewidth]{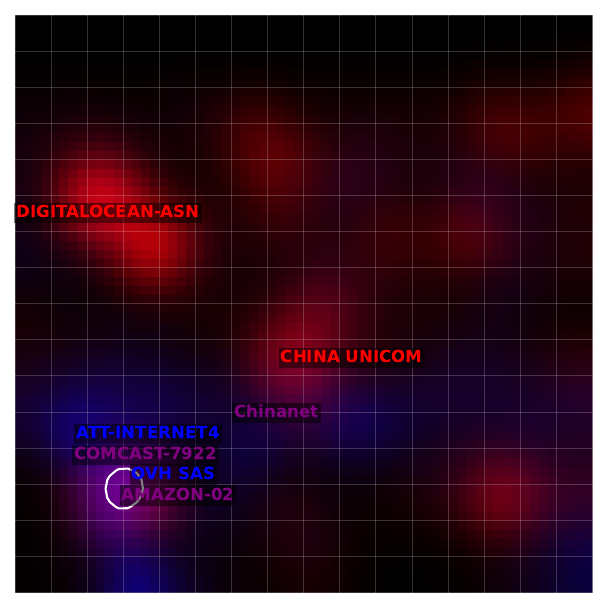}
    \caption{Dual-density plot overlaying adjusted attack densities (red) and Ethereum node densities (blue) in IPv4 space. Areas of overlap appear purple, with an 80\% density overlap contour line, indicates IP space where attacks originate and nodes co-occur.}
    \label{fig:dual-density}
\end{figure}

\cref{fig:dual-density} shows a dual-density plot overlaying the adjusted attack densities and Ethereum node location densities across IPv4 space, using Gaussian smoothing ($\sigma=3.5$) to aid visualization. The plot reveals limited overlap between regions of high attack sources and high concentration of Ethereum nodes, suggesting that attacks are not predominantly originating from networks where Ethereum nodes are hosted. Quantitative analysis supports this: the Jaccard Index—measuring the overlap between ASNs in the top 20\% for both adjusted attack counts and node counts—is low at 0.274. This indicates that only about 27\% of ASNs are common between high attack activity and high node presence. Additionally, the substantial Kullback-Leibler Divergence of 6.439 and the large Earth Mover's Distance of 1,265.287 between the distributions of adjusted attack counts and node counts per ASN indicate significant divergence. These findings support our hypothesis that attacks primarily originate from different networks than those hosting Ethereum nodes, highlighting targeted attack patterns against the networks where Ethereum nodes are concentrated.

\subsection{Scanning Data}

To understand the exposure of all Ethereum nodes to the attacks we observed on our honeypots, we conducted reconnaissance and empirical analysis from an adversary's perspective. Using a web scraper on a popular node explorer service that regularly crawls the mainnet, we identified 6,185 active beacon nodes recently observed online. We then conducted TCP port scans against these nodes, targeting both the top 100 ports from our honeypot data and default RPC/P2P ports.

Ports are labeled as either \texttt{closed}, \texttt{filtered}, or \texttt{open}. Closed ports are accessible, but have no listening application; filtered ports have indeterminate status, likely due to firewall rules or packet filtering; open ports are actively accepting connections. Even if a host currently has no exploitable vulnerabilities on its open ports, their exposure to the Internet makes it vulnerable to future exploits. The risk extends from the P2P network to the broader Internet, as compromised P2P nodes could serve as unwitting platforms for future attacks. 

25.68\% (1588/6184) of nodes scanned had at least one open port beyond their P2P ports. \cref{fig:open-ports} shows the top 50 open ports scanned. 7.45\% had a single port open, while 18.22\% had multiple open ports. The full distribution of open ports detected is in \cref{fig:open-ports} in the Appendix. 

\cref{tab:port-table} presents the most prevalent individual ports and port combinations observed. SSH (port 22) was the most common, present on 13.63\% of nodes, followed closely by HTTPS (443) and HTTP (80) at approximately 12.7\% each. Notably, 2.99\% of nodes exposed the Ethereum RPC port (8545), and 2.36\% had port 3000 accessible, which is default port for the open-source monitoring tool, Grafana.

The most frequent port combinations suggest deliberate administrative configurations. The most common combination was HTTP/HTTPS (80, 443) at 4.66\%, followed by a HTTP/HTTPS with SSH access (22, 80, 443) at 1.57\%. Of particular concern, 1.12\% of nodes exposed both SSH and RPC ports (22, 8545). A significant finding was that 213 nodes (3.44\% of the total sample) had default RPC ports (8545 or 8546) exposed to the Internet. %

To validate our findings and exclude potential honeypots, we examined port distribution patterns. The absence of large ranges of open ports suggests catch-all honeypots were not present in our dataset. Furthermore, our P2P session data confirmed that we established productive peer connections with all nodes showing open ports, affirming their status as legitimate network participants rather than passive monitoring systems.

The concentration of open ports around common administrative services (SSH, HTTP/S) and monitoring tools (Grafana) indicates that node operators are primarily exposing ports for management purposes. However, the presence of exposed RPC ports and various service combinations suggest some operators may prioritize convenience over security, potentially creating vulnerabilities exploitable by future attacks.

\begin{table}[t]
\centering
\caption{Generalized Linear Mixed Model results.}
\small
\begin{tabularx}{\columnwidth}{lXXXX}
\hline
 & Coeff. & SE & z & P>|z| \\
\hline
\textbf{Fixed Effects} & & & & \\
Intercept & 2.804 & 0.100 & 28.117 & <0.001 \\
Group (Experimental) & 1.111 & 0.008 & 147.289 & <0.001 \\
\hline
\textbf{Random Effects} & & & & \\
Group Variance (Region) & 0.050 & 0.030 & & \\
\hline
\textbf{Model Fit Statistics} & & & & \\
Obs & 100{,}000 & & & \\
Regions & 5 & & & \\
\hline
\end{tabularx}
\label{tab:glmm_results}
\end{table}

\begin{table}[t]
\centering
\caption{Top 5 most common open ports and port combinations.}
\small
\begin{tabular}{lrr|lrr}
\toprule
\textbf{Port} & \textbf{\%} & \textbf{Count} & \textbf{Port Combination} & \textbf{\%} & \textbf{Count} \\
\midrule
22       & 13.63\% & 843 & (80, 443)          & 4.66\%  & 288 \\
443      & 12.75\% & 788 & (22, 80, 443)      & 1.57\%  & 97  \\
80       & 12.73\% & 787 & (22, 8545)         & 1.12\%  & 69  \\
8545     & 2.99\%  & 185 & (22, 3000)         & 0.87\%  & 54  \\
3000     & 2.36\%  & 146 & (22, 80)           & 0.65\%  & 40  \\
\bottomrule
\end{tabular}
\label{tab:port-table}
\end{table}

\section{Discussion}

This work addresses a knowledge gap in cross-network threat intelligence, where nodes in P2P overlay networks can be specifically targeted as host platforms for broader attacks against applications and infrastructure on the broader internet. We chose the Ethereum mainnet as our initial network to test this hypothesis due to its scale, availability of measurement utilities, and emergence as a \textit{de facto} global standard for public blockchain infrastructure. Our findings provide initial evidence supporting our hypothesis, as our nodes were exposed to higher-than-expected reconnaissance and targeted exploitation attempts. Our results contribute to the broader internet measurement community by demonstrating a replicable approach to collecting and analyzing attack data in a large-scale P2P overlay networks. 

\textbf{Implications for Decentralized Overlay Networks.} The implications of our findings extend beyond Ethereum to other P2P overlay networks. %
As these networks continue to gain relevance, %
the lack of comprehensive and recurring threat assessments limits our ability to identify and mitigate emerging risks. Networks under the web3 umbrella prioritize permissionlessness and transparency, which inadvertently simplifies the kill chain for would-be attackers. Our study highlights the need for standardized threat assessments across diverse P2P networks to identify unique attack vectors and adapt appropriate mitigation measures for each. \textit{The open architectures and reachability of public P2P overlay networks are both a core strength and an attack surface.}

\textbf{Attack-Induced Impacts.} A secondary contribution of our work is a measurement-based analysis of how active attacks affect Ethereum nodes' performance. Open ports alone can cause tangible damage, limiting a node's normal functionality. By running honeypots on our nodes, we demonstrate potential damage: when a node is used in a reflection-based attack, or is aggressively scanned, our data show significantly increased CPU loads. This resource demand forces the kernel to de-allocate resources from user processes. As visualized in~\cref{fig:sip-attack}, during an attack recorded by the Sentrypeer~\cite{sentrypeer} honeypot, the affected node lost 20\% of its peers and was unable to perform Ethereum beacon node functions—effectively a denial-of-service.

\subsection{Limitations and Future Work}

Our study has limitations that call for future research. Employing higher-interactivity honeypots tailored for P2P environments could enhance data quality and offer deeper insights into targeted behaviors. Additionally, although our methodology sought a balance between depth and global coverage of a targeted network to gather this initial evidence, future work should employ a broader array of vantage points spanning multiple P2P overlay networks to fully assess cross-network threats. Focusing on Ethereum due to its prominence, we intend to extend this study to other overlay networks to determine if these attack patterns are consistent across different contexts. Validating findings in other prominent P2P networks will help confirm the generalizability of our results and is a direction for future work. %

\section{Conclusion}

We examined active reconnaissance attacks on Ethereum P2P network nodes by deploying honeypots alongside actual Ethereum nodes across globally distributed VPs. We discovered that these nodes not only faced an increased number of attacks but also specific types targeting particular ports and services. By performing port scans on reachable nodes, we provided evidence that the threat landscape observed on Ethereum nodes is applicable to the broader P2P communities. Our findings offer initial insights into potential mitigation strategies aimed at enhancing the security of P2P overlay networks.

\bibliographystyle{ACM-Reference-Format}
\bibliography{refs}

\appendix
\section{Ethics} To mitigate any potential privacy issues in our collection and sharing of this threat intelligence, we followed best practices for measurement studies~\cite{pauley2023understanding}. We refrained from disclosing specific IP addresses and implemented anonymization when presenting attack patterns and performed data encryption in both transit and at rest. Additionally, we limited our scanning methods to publicly accessible information and employed non-intrusive, narrowly-scoped port scans to ensure minimal disruption to any given node. Future studies should continue to consider these ethical dimensions, especially as the data collected could inadvertently reveal operational practices of node operators that could put them at greater risk. 

\section{Recommendations for Operators and Protocol Researchers}

The detailed insight into network-level distribution of attack sources presented in this study can inform operators in deploying proactive measures. Specifically, operators could throttle or block IP spaces associated with high scan rates or attack activity but low peer density, mitigating potential risks without compromising access to legitimate peers. Our research suggests that by leveraging this data, node operators can implement IP filtering measures and other defensive tactics tailored to known threat vectors.

Another potential security avenue is the use of private peering arrangements or mesh VPNs. For private peering arrangements, node operators could make sure that the core of their peer list is other known actors that have met certain reputation-based criteria\footnote{For an ongoing discussion on such proposed solutions, see~\cite{proof_of_validator}.}. An even more secure network security approach would be to establish a large, decentralized, mesh VPN for peer connectivity, where the nodes can granularity control their exposure to the public Internet. This could serve as a necessary mechanism to protect the health of a network that has significant economic value and public dependence.

Our findings suggest that implementing certain network-layer security enhancements does not necessarily have to infringe upon a network’s decentralized or permissionless properties. Selective security measures may help secure critical infrastructure within these networks while preserving their openness and accessibility for new node operators.

\section{Supplementary Results}

\begin{figure}[H]
    \centering
    \includegraphics[width=\linewidth]{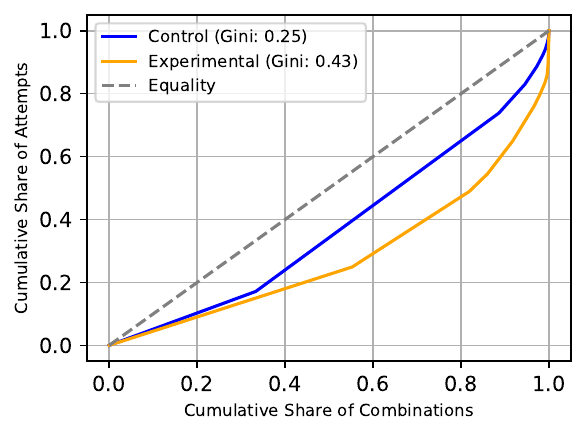}
    \caption{Lorenz curve showing the greater concentration of username:password login attempts against nodes.}
    \label{fig:lorenz}
\end{figure}

\begin{figure}[H]
    \centering
    \includegraphics[width=\linewidth]{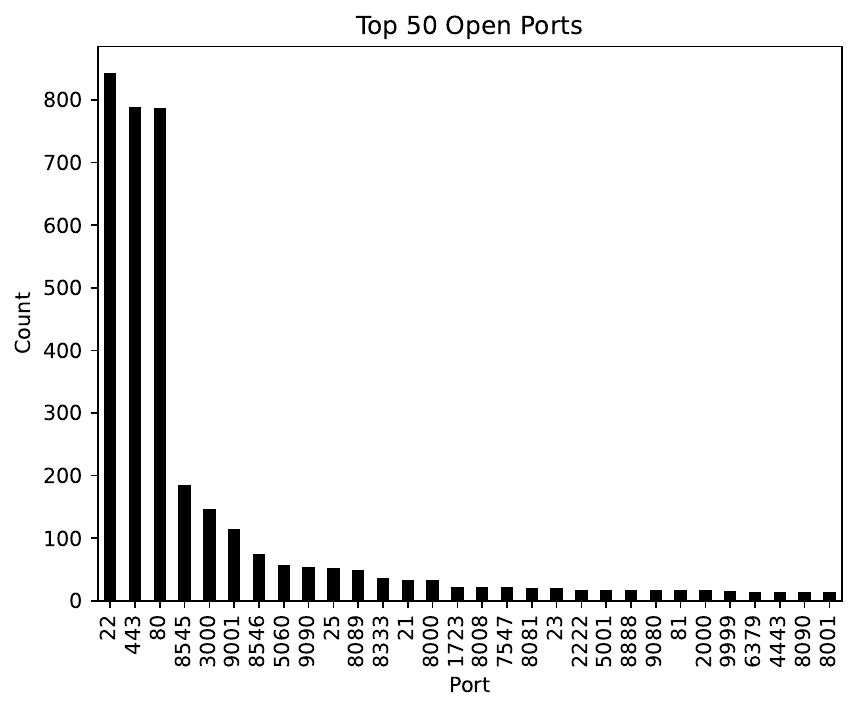}
    \caption{The top 50 open ports by active scanning of online Ethereum beacon nodes excluding P2P ports.}
    \label{fig:open-ports}
\end{figure}

\begin{figure}[H]
    \centering
    \includegraphics[width=\linewidth]{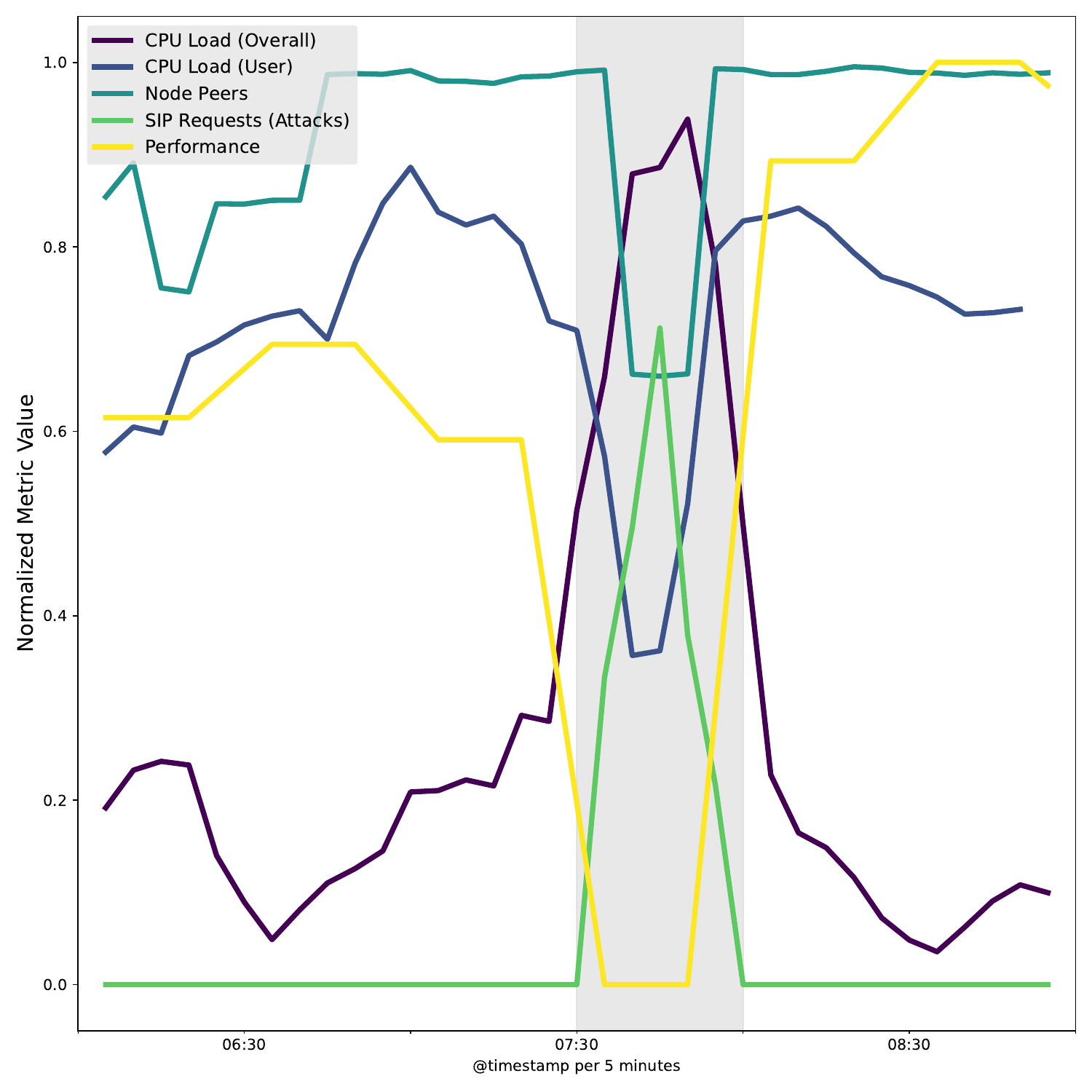}
    \caption{Example of an active attack against the Sentrypeer honeypot on an experimental server, illustrating that during the attack, peer count dropped as the attack rate increased. Overall CPU load increases, decreasing the CPU cycles allocated to user processes, dropping attestation Performance. The node recovers post attack. Shown in 5 min ticks with moving average.}
    \label{fig:sip-attack}
\end{figure}

\begin{table}[H]
\centering
\caption{Honeypot containers with service and port coverage.}
\label{tab:honeypot-containers} \small
\begin{tabularx}{\columnwidth}{|p{1.9cm}|p{1.5cm}|X|X|}
\hline
\textbf{Honeypot} & \textbf{Interaction} & \textbf{Target\linebreak Services} & \textbf{Ports} \\ \hline
Adbhoney & Medium & Android Debug Bridge (ADB) & 5555/TCP \\ \hline
CiscoASA & Medium & Cisco ASA VPN & 5000/UDP, 8443/TCP \\ \hline
Citrixhoneypot & Medium & Citrix Gateway & 443/TCP \\ \hline
Conpot & Low & Guardian AST, IEC 104, IPMI, \newline Kamstrup 382 & 10001/TCP, 2404/TCP, \newline 161/UDP, 623/UDP \\ \hline
Cowrie & High & SSH, Telnet & 22-23/TCP \\ \hline
Ddospot & Low & DDoS-related services & 19/UDP, 53/UDP, \newline 123/UDP \\ \hline
Dicompot & Low & DICOM Medical Imaging & 11112/TCP \\ \hline
Dionaea & High & FTP, SMB, HTTP, MySQL, \newline and more & 20-21/TCP, 445/TCP, \newline 1433/TCP, 3306/TCP \\ \hline
Elasticpot & Low & Elasticsearch & 9200/TCP \\ \hline
Heralding & Medium & IMAP, SMTP, PostgreSQL & 110/TCP, 5432/TCP, \newline 993/TCP \\ \hline
Honeytrap & Medium & Various protocols & All other ports \newline not covered \\ \hline
IPPhoney & Low & Internet Printing Protocol (IPP) & 631/TCP \\ \hline
Mailoney & Medium & SMTP, Mail Relay & 25/TCP, 587/TCP \\ \hline
Medpot & Low & Medical Device Communication (HL7) & 2575/TCP \\ \hline
Redishoneypot & Low & Redis & 6379/TCP \\ \hline
Sentrypeer & Medium & SIP & 5060/TCP, 5060/UDP \\ \hline
Snare & Low & HTTP & 80/TCP \\ \hline
Spiderfoot & Medium & OSINT, Reconnaissance & 8080/TCP \\ \hline
Tanner & Medium & API, PHP & Various \\ \hline
Wordpot & Low & WordPress Vulnerabilities & 8080/TCP \\ \hline
\end{tabularx}
\end{table}

\end{document}